\documentclass[doublecol]{epl2}
\usepackage{amsmath}

\shorttitle{Nucleation of bubbles} 
\institute{Center for Nonlinear Phenomena and Complex Systems CP 231,
  Universit\'{e} Libre de Bruxelles, Blvd. du Triomphe, 1050 Brussels, Belgium
}
\abstract{The nucleation of vapor bubbles within a superheated fluid is studied using density functional theory. The nudged elastic band technique is used to find the minimum energy pathway from the metastable uniform liquid to the stable uniform gas thus emphasizing the analogy between the the nucleation problem and that of chemical reactions. The result is both an accurate determination of the critical nucleus and an unbiased description of the density profile at various points along the path between the free energy extrema. This calculation is compared to two other methods: the use of parametrized profiles and constrained minimization of the free energy. The results indicate that the recent claim, based on the constraint method, that bubble nucleation and growth involves an activated instability is incorrect.}

\begin{document}

\title{Theoretical description of the nucleation of vapor bubbles in a
superheated fluid}
\author{James F. Lutsko \thanks{\email{Email: jlutsko@ulb.ac.be}}}
\maketitle

\section{Introduction}

The process of nucleation, e.g. of solids from supercooled liquids or of
liquids from supersaturated gases, has become the subject of intense
interest in recent years driven by applications in physics, chemistry and
biology. Practical problems such as the control of protein crystallization%
\cite{VekilovCGDReview2004} have
motivated experimental, theoretical and computational research which has
uncovered new fundamental physics such as the important role that
intermediate metastable phases may play in nucleating new phases\cite%
{VekilovCGDReview2004, tWF, OxtobyProtein, GuntonProtein, Lutsko_PRL_2006}.
In cases where this is important, nucleation becomes more complex than the
simple picture of barrier crossing that underlies classical nucleation
theory (CNT). Indeed, there is some theoretical evidence that the nucleation
of solids from gases (a simple model of the precipitation of solids from
solution) may be nonclassical even in the case of simple fluids\cite%
{Lutsko_PRL_2006}. Recently, it has even been claimed that the nucleation of
gas bubbles in a superheated fluid may be nonclassical due to the presence
of an ``activated instability''\cite{Corti_PRL}.

Density functional theory (DFT) provides a natural approach to the
description of nucleation. DFT is based on the fact\cite%
{MerminDFT,HansenMcdonald} that the free energy is a unique functional of
the local density, $\rho(\vect{r})$. In simple systems, different phases
correspond to different local densities: for example, in gases and liquids,
the density is a constant $\rho(\vect{r}) = \bar{\rho}$ whereas in a solid
it varies over the length scale of the lattice. Nucleation can then be
viewed as a process of moving from one point in density-function space to
another via the most probable path. This is analogous to the usual
description of a reaction pathway in a chemical system. As is usually done
in chemical problems, it is natural to assume that the most probably path
will be the minimum free-energy path (MFEP) so that the problem reduces to
one of finding the MFEP connecting the initial and final states given the
model free energy functional. 

The local density is a field so that in principle its value at every point
in space can vary independently giving it an infinite number of degrees of
freedom. Here, attention will be restricted to spherically symmetric
profiles which will furthermore be discretized on a lattice giving a
finite, but still large, number of degrees of freedom. The free energy is therefore a
surface in this high dimensional space making its characterization
difficult. This is analogous to the problem of characterizing reaction
paths in chemical systems. Fortunately, recent years have seen the development of
several sophisticated methods for addressing the problem of finding
optimal pathways on high-dimensional surfaces(i.e., eigenvalue
following, string method, elastic band methods ...)\cite{Wales}. One of the
primary goals of this paper is to show that these methods can be applied to
the description of nucleation.

Liquid-vapor nucleation in simple fluids has, of course, been studied
previously and might seem too classical to warrant further effort. Although previous work has mostly focused on nucleation of 
droplets in a supersaturated gas, the issues are the same for droplet and bubble nucleation. Early studies\cite{Lee, Oxtoby_Evans} focused on 
the determination of the critical cluster since this is accessible (it is an extremum of the free energy) and determines the free 
energy barrier and, hence, the nucleation rate. In the 1990's Reiss et al\cite{Reiss1, Reiss2, Reiss3} developed an approximate simulation technique 
for studying non-critical clusters and this led to parallel theoretical developments\cite{talanquer:5190} which have recently been extended 
to the study of gas bubbles in superheated fluids\cite{Corti_PRL}. (This is referred to as the ``constraint'' method below.) For use in simulations, 
this  method assumes a decoupling of clusters from the surrounding bulk. More recently, 
importance-sampling techniques for Monte Carlo simulations have been developed which do not depend on assumptions like decoupling\cite{frenkel_gas_liquid_nucleation,FrenkelSmit}. These developments in simulation techniques are a primary motivation for the present work which aims to develop a similarly unbiased theoretical description of nucleation.

In this paper, three methods will be used to determine the properties of
bubble nucleation. The simplest is the use of a parametrized profile. It is
assumed that the density profile of a bubble is sigmoidal and is
approximated by a generalization of a hyperbolic tangent. These profiles
have two important parameters: the location of the interface, i.e. the bubble
radius, and the width of the interface. The approximate pathway is
constructed by minimizing the free energy with respect to the width of the
interface for different values of the radius.  The second method is the constrained optimization
method originated by Talanquer and Oxtoby\cite{talanquer:5190} for the study
of droplet nucleation and adopted by Uline and Corti (UC)\cite{Corti_PRL} for bubble nucleation where the free
energy is minimized while subject to a constraint that enforces a particular
bubble geometry. The third method is the
nudged elastic band (NEB)\cite{NEB} which is a chain of states method for
constructing the MFEP. It will be shown that of the three methods, the NEB
provides the most natural and reliable determination of the reaction path
and that the instability identified by UC is an artifact of their method.

\section{DFT calculations}

In DFT, the grand potential is written as 
\begin{equation}
\beta\Omega\left[ \rho\right] =\beta F\left[ \rho\right] -\beta\mu\int
\rho\left( \vect{r}\right) d\vect{r}+\int\beta\phi\left( \vect{r}\right)
\rho\left( \vect{r}\right) d\vect{r}  \label{1}
\end{equation}
where $\beta=1/k_{B}T$, $\mu$ is the chemical potential, $\phi\left( \vect{r}%
\right) $ is an external one-body field and $\rho\left( \vect{r}\right) $ is
the local density. The functional $\beta F[\rho]$ is related to the
Helmholtz free energy\cite{HansenMcdonald,EvansDFT}. To be precise, DFT
tells us that for fixed $T$ , $\mu$ and $\phi\left( \vect{r}\right) $, the
equilibrium density is a stationary point of the functional $\beta\Omega%
\left[ \rho\right] $ . In this sense, there is little direct relevance of
this functional to the problem of nucleation since nucleation typically
occurs at some given fixed field (either no field or the gravitational
field) so that this functional can only be used to describe a few simple
states:\ the bulk liquid and vapor and, potentially, the critical nucleus.
All other intermediate states will, in general, not extremize the
functional. However, it is usually assumed that at fixed field, $\beta\Omega%
\left[ \rho\right]$ plays the role of a potential and that it can therefore
be used to deduce interesting quantities such as the energy barrier for
transition from one state to another.

The calculations reported here will be based on a recently developed DFT
model consisting of a modified Fundamental Measure model for the
short-ranged repulsion and a Van der Waals tail for the long-ranged
attraction\cite{lutsko_jcp_2008_2}. This model gives a quantitatively
accurate description of the inhomogeneous Lennard-Jones liquid including the
planar liquid-vapor interface, the liquid near a wall and the liquid
confined to a slit pore. The bulk thermodynamics was approximated using the
empirical 33-parameter equation of state of Johnson, Zollweg and Gubbins 
\cite{JZG}.

Three methods are used to explore the free energy landscape. The
first and simplest is the use of a parametrized profile. The
parametrization used here is a modified hypertangent, 
\begin{equation}  \label{param}
\rho\left( r\right)=\rho _{l}+\left( \rho _{v} -\rho _{l}\right) \frac{1+br}{%
1+\left( br\right) ^{2}}\frac{1-\tanh \left( A(r-R)\right) }{1-\tanh \left(
-A_{}R\right) }
\end{equation}%
with%
\begin{equation}
b=\frac{2A}{e^{2AR}+1}.
\end{equation}
and where $\rho_l$ ($\rho_v$) is the density of the bulk liquid
(vapor) for the given temperature and chemical potential.
This function is constructed so that  it approaches $\rho_l$ as $1/r$ at
large distances while also satisfying $d\rho/dr \rightarrow 0$ as $r
\rightarrow 0$. In the calculations presented below, the free energy is
minimized with respect to the width parameter $A$ for fixed values of the
radius, $R$.

The second method used was the constraint method of Talanquer and Oxtoby\cite%
{talanquer:5190} as adapted by Uline and Corti\cite{Corti_PRL}. Here, one
characterizes a bubble by some property of the density which can be written
schematically as $g\left( \left[ \rho \right] ,\Gamma \right) =0$ where the
notation indicates that the left hand side is a functional of the local
density and that it also depends on one or more parameters denoted
collectively as $\Gamma $. Different values of the parameters correspond to
bubbles having different properties (mass, radius, ...). It is assumed that
for fixed parameters, the most energetically accessible structure is that
which minimizes the free energy subject to the structural constraint. This
is implemented via a Lagrange multiplier by extremizing the functional $%
\beta \Omega \left[ \rho \right] -\alpha g\left( \left[ \rho \right] ,\Gamma
\right) $ giving the equations%
\begin{equation} \label{Lagrange}
\frac{\delta \beta F\left[ \rho \right] }{\delta \rho \left( \vect{r}\right) 
}-\beta \mu -\alpha \frac{\delta g\left( \left[ \rho \right] ,\Gamma \right) 
}{\delta \rho \left( \vect{r}\right) } =0   \\
\end{equation}%
and $g\left( \left[ \rho \right] ,\Gamma \right) =0$. In this picture, the
role of the constraint is to reduce the number of degrees of freedom from an
essentially infinite number needed to describe all possible density
functionals to a few, the collection of parameters $\Gamma $, which then
define a path from the uniform liquid to the uniform vapor. The question
then is what to use for the structural constraint. First, note that if there
is no structural constraint, then the uniform state satisfies the usual
thermodynamic relation $\frac{\partial Vf\left( \rho \right) }{\partial \rho 
}=\mu $,where $f\left( \rho \right) $ is the bulk free energy per unit
volume, and there are two solutions corresponding to the uniform vapor, $%
\rho _{v}$, and liquid, $\rho _{l},$ states. Perhaps the most natural choice
for the constraint is the number of atoms "missing" from the bubble relative
to the liquid background,%
\begin{equation}  \label{simple}
g\left( \left[ \rho \right] ,\Gamma \right) =\int \left( \rho _{l}-\rho
\left( \vect{r}\right)\right) d\vect{r}- \Delta N  
\end{equation}%
where $\Delta N$ is the desired deficit. In this case, the only parameter is 
$\Gamma =\Delta N$ . However, substitution into Eq.(\ref{Lagrange}) shows that this
changes the chemical potential from $\mu $ to $\mu +\alpha $ so that the
liquid density far from the bubble would necessarily be incorrect. This
seems to indicate that a localized bubble can only be described by a
constraint that vanishes at large $r$ . UC therefore propose%
\begin{equation}
g_{UC}\left( \left[ \rho \right] ,\Gamma \right) =\int \Theta \left( \lambda
-r\right) \rho \left( r\right) d\vect{r}-N
\end{equation}%
where $\Theta(x)$ is the step function and there are two parameters, the
"radius" $\lambda $ and the total number of atoms in the bubble, $N$\cite{Corti_PRL}. (For
the reverse problem of droplet nucleation, Talanquer and Oxtoby also use
this constraint but they say they adjust the bulk outside the volume in a
way which appears to imply the use of Eq.(\ref{simple})\cite{talanquer:5190}.) UC show that, for
a fixed value of $N$, solutions only exist for $\lambda $ less than some
critical value and this is interpreted as indicating that bubbles of larger
size are unstable. I propose here another constraint that is based on
defining the size of a bubble implicitly as the position at which the density
becomes greater than some value, $\rho _{\ast }$, 

\begin{equation}  \label{constraint}
g_{\ast }\left( \left[ \rho \right] ,\Gamma \right) =\int \Theta \left( \rho_{\ast } - \rho
\left( r\right)\right) d\vect{r}-\Gamma .
\end{equation}%

This is physically appealing and is also similar to a criterion used to
define clusters in some computer simulations\cite%
{frenkel_gas_liquid_nucleation}. In this case there are in principle two
parameters: $\Gamma $, which is a measure of the volume of the bubble, and $%
\rho_*$ which will be taken to be $\frac{1}{2}%
\left(\rho _{l}+\rho _{v}\right) $ .

Finally, the third method used is the nudged elastic band\cite{NEB}. These
calculations begin with some number, $N$, of  density profiles corresponding to Eq.(\ref{param}%
) with values of $R$ spaced evenly from $R=0$ to some $R_{max}$. The various
profiles represent points along a path from the initial state (the uniform
fluid corresponding to $R=0$ in Eq.(\ref{param})) to the final state which
here is taken to be again a sigmoidal with a fixed, large radius and
optimized width. (These endpoints remain fixed during the calculation and
the remaining images are referred to as ``moveable''.) Unlike the first
method, each of these images are not restricted to be given by the
parametrized form of Eq.(\ref{param}) but is instead a discretized density
profile, $\rho(r_i)$, that can assume any shape. If the sum of the free
energies of all of the moveable profiles was minimized, the final result
would be that each profile would tend toward one of the minima - either the
uniform liquid or the uniform vapor, depending on the starting point. To
avoid this, the NEB involves two elements. First is the addition of
artificial elastic forces between neighboring profiles so that the images
are forced to be more or less uniformly distributed along the MFEP. To
define the elastic forces, the distance between two profiles was defined to
be 
\begin{equation}
\left|\rho_1-\rho_2\right|^{2}=\int w(r) \left(\rho_1(\vect{r})-\rho_2(%
\vect{r})\right)^2 d\vect{r}.
\end{equation}
The definition allows for a weight, $w(r)$, since there are at least two
natural choices one could make in a spherical geometry. The first is, of
course, $w(r)=1$ while the second is $w(r)=r^{-2}$. For spherically
symmetric density distributions, the first choice has the effect of giving
more weight to differences in density at large $r$ than at small distances
while the second weights all radial points equally. Calculations were
performed using both definitions and it was found, as shown below, that the
choice had no effect on the final results except for affecting the relative
spacing of the images along the MFEP: the second choice gives more images at
small $r$ than the first choice but the path so defined was virtually
identical.

The second element of the NEB method is that this combination of the free
energies and artificial elastic forces is minimized under the constraint
that the elastic forces only act along the MFEP and the force arising from
the free energies act only perpendicular to the MFEP (the ``nudging''). This
method is not guaranteed to give the exact MFEP but it has proven a useful
heuristic in numerous applications. A further modification involves choosing
one image and forcing it to climb the gradient along the MFEP without any
spring forces\cite{NEB-CI}. This gives a very accurate and convenient
characterization of the saddle point. The relaxation of the system of images and harmonic forces
was performed using the the fast inertial relaxation engine
algorithm\cite{FIRE}.

\section{Results}

The three methods were compared for the case of a Lennard-Jones potential, $%
V(r)=4\epsilon \left( \left( \frac{\sigma}{r} \right)^{12} - \left( \frac{%
\sigma}{r} \right)^{6} \right)$ with a cutoff at $r=4\sigma$. The
calculations reported here were performed at a reduced temperature $%
k_{B}T/\epsilon = 0.8$ and the chemical potential was fixed at $10\%$ below
coexistence corresponding, roughly, to the state conditions used by UC\cite%
{Corti_PRL}. The densities of the coexisting liquid and vapor were found to
be $\rho_{l}\sigma{^3} = 0.784$ and $\rho_v\sigma^3 = 0.008$ respectively.

\begin{figure}
\centerline{\includegraphics[height = 7cm, angle=-90]{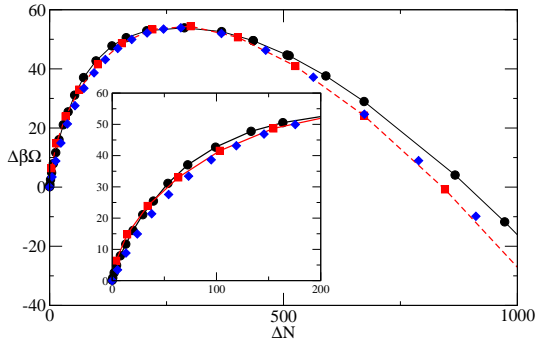}}
\caption{(Color online) Excess free energy as a function of the atomic
deficit as calculated using DFT with the proposed structural constraint
given in Eq.(\protect\ref{constraint}), circles and solid line, using a
modified hyperbolic tangent profile, Eq.(\protect\ref{param}), squares and
broken lines, and the NEB method, diamonds. }
\label{fig1}
\end{figure}

Figure \ref{fig1} shows the excess free energy, calculated using all three methods, as a function of
the size of the bubble as quantified by the number deficit (i.e. as defined via eq.(\ref{simple})). The constraint method
has been implemented using eq.(\ref{constraint}). All three
methods are in close agreement for the height of the free energy barrier and
the size of the critical cluster. However, away from the critical cluster,
the constraint method gives systematically higher free energies than do the
other two methods. For the initial part of the bubble formation, from the
uniform fluid to the critical cluster, the NEB method gives the lowest free
energies and, hence, is closest to the MFEP. For bubbles larger than the
critical cluster, the NEB energies are very close to, but slightly higher
than, the those calculated using the parametrized profile. This is probably
due to the fact that the end point of the chain of states used in the NEB
calculation is not the uniform gas - since the uniform gas is a bubble of
infinite size, it cannot serve as the end of the chain - but, rather, a
parametrized profile which thus distorts the NEB chain away from the MFEP.

\begin{figure}
\centerline{\includegraphics[height = 7cm, angle=-90]{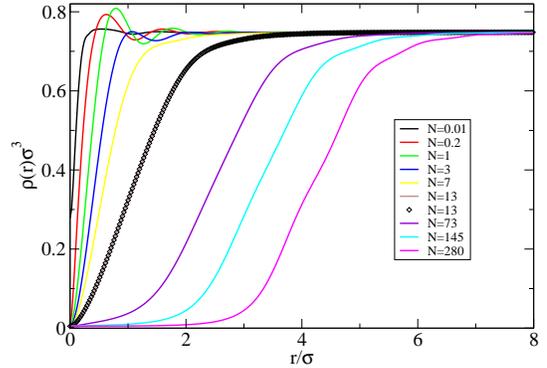}}
\caption{(Color online) Structure of small bubbles along the MFEP as
determined using the NEB method. Structures up to $N=13$ were calculated
using the metric with $w(r)=r^{-2}$ weight whereas structures for $N=13$ and
larger were performed including the $w(r)=1$. The $N=13$ structure is shown
as calculated both ways (line is without the weight and the symbols are with
the weight) illustrating the insensitivity of the calculation to the weight
used in the metric. }
\label{fig2}
\end{figure}

Figure \ref{fig2} shows the density profiles calculated using the NEB
method. The bubble initially forms as a small drop in the density very close
to the origin. The central density quickly becomes that of the coexisting
gas but the bubble is still very small due to the fact that the interface is
extremely narrow. At this stage, some structure forms in the liquid which is
similar to that seen in a fluid near a hard wall\cite{lutsko_jcp_2008_2}.
The bubble grows via the broadening of the interface and as the interface
broadens, the liquid structure vanishes. Only when the interface is
sufficiently broad, in this case, when the number deficit is about 70 atoms
or so, does a recognizable sigmoidal structure develop. Further growth
consists of a displacement of the interface with little change in structure.

\begin{figure}
\centerline{\includegraphics[height = 7cm, angle=-90]{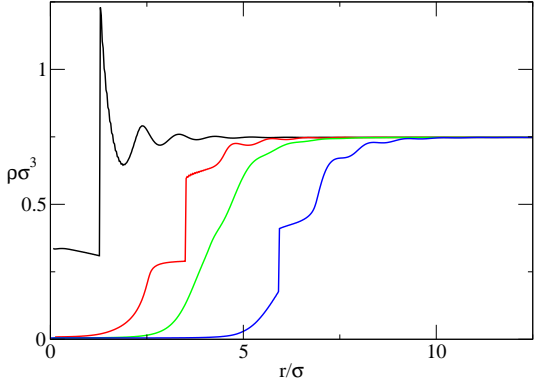}}
\caption{(Color online) Density profiles for several different bubble sizes
of, from left to right, $\Delta N = 4, 163, 288$ and $866$ as calculated
using the constraint method.}
\label{fig3}
\end{figure}

When the free energy is minimized for a spherically symmetric density
profile with the UC constraint, the results of UC are reproduced: namely,
for fixed $N$, there is a maximum value of the radial parameter, $\lambda$,
beyond which no stable profile exists. The physics of this apparent
instability will be the subject of the next section.

Using the proposed constraint, a different picture emerges. It is possible
to stabilize the density profile for all values of the parameter $\Gamma$.
For each value of $\Gamma$, the total atomic deficit can be calculated and
the resulting free energy curve is shown in Fig. 1 which confirms the CNT
picture and shows no sign of an instability. However, because the constraint has the effect of shifting the chemical potential inside the bubble, but not outside, (see eq.(\ref{Lagrange})), the profiles generated are discontinuous, as can be seen in Fig. \ref{fig3}. The only continuous profile obtained is for the special case of the critical nucleus, for which the Lagrange multiplier $\alpha$ is zero. The discontinuity of the profiles is responsible, at least in part, for the higher free energies calculated for this model as shown in Fig. \ref{fig1}. 


\section{Illustration on a simple model}

It is useful to consider
a simple, but instructive, analytic model for which it is possible to see why the constraint method can give instabilities. 
Consider the square-gradient model first introduced
by van der Waals\cite{VDW2,VDW1},%
\begin{equation}
\beta F\left[ \rho \right] =\int \left[ \beta f\left( \rho \left( r\right)
\right) +\frac{1}{2}\tau \left( \frac{\partial \rho \left( r\right) }{%
\partial r}\right) ^{2}\right] d\vect{r}
\end{equation}%
where again $f\left( \rho \right) $ is the free energy per unit volume of
the uniform fluid and where the constant $\tau $ is called the accommodation
factor. This model is not quantitatively very accurate, but it is a
systematic approximation to more realistic models and is still often used in
the literature. Combined with the structural constraint, this can be used to
determine the density. The problem is simplified if one uses a parametrized
density and here I consider a very simple piecewise linear approximation,%
\begin{equation}
\rho \left( r\right) =\left\{ 
\begin{array}{c}
\rho _{0},\;r<R \\ 
\rho _{0}+\frac{\rho _{\infty }-\rho _{0}}{w}\left( r-R\right) ,\;R<r<R+w \\ 
\rho _{\infty },\;R<r%
\end{array}%
\right.   \label{linear}
\end{equation}%
Potentially, the width, $w$, the bubble density, $\rho _{0}$ , the outer
density, $\rho _{\infty }$ and radius $R$ are determined by minimizing the
grand potential (with the structural constraint). For the outer density,
this gives $\rho _{\infty }=\rho _{l}$. However, for the remaining
parameters, this is still too complicated, so I consider the simplest case
that the width goes to zero. If this limit is taken, while holding $\gamma
\equiv \tau /w$ constant, then the resulting functional using the UC
constraint is easily shown to be
\begin{gather}
\beta \Omega \left[ \rho \right] -\alpha g_{UC}\left( \left[ \rho \right]
,\Gamma \right) \label{long} \\
=\frac{4\pi }{3}R^{3}\left\{ f\left( \rho _{0}\right)
-f\left( \rho _{l}\right) -\mu \left( \rho _{0}-\rho _{l}\right) \right\} 
\notag \\
+4\pi \gamma R^{2}\left( \rho _{l}-\rho _{0}\right) ^{2}  \nonumber \\
-\alpha \left\{ 
\begin{array}{c}
\left(\frac{4\pi }{3}\lambda ^{3}\rho _{0}-N\right) \Theta(R-\lambda) \\ 
+\left(\frac{4\pi }{3}R^{3}\rho _{0}+\frac{4\pi }{3}\left( \lambda
^{3}-R^{3}\right) \rho _{l}-N\right)\Theta(\lambda-R)
\end{array}
\right\}   \nonumber
\end{gather}%
Notice that in the absence of the constraint, $\alpha =0$, this is the usual
functional assumed in classical nucleation theory if one takes $\rho
_{0}=\rho _{v}$ . In that case, the free energy is a function of a single
parameter, the radius $R$, and it is easy to see that $\beta
\Omega \left[ \rho \right] $ as a function of $R$ shows a minimum at $R=0$
(the liquid) a maximum at some finite value of $R$ and a monotonic decrease
for larger values of $R$. This is just the CNT picture of nucleation as a
process of thermally activated barrier crossing. Using the proposed
constraint, the functional to be minimized is the same as the UC functional
except the Lagrange multiplier term is $-\alpha \left( \frac{4\pi }{3}%
R^{3}-\Gamma \right) $, provided that $\rho _{0}<\rho _{\ast }<\rho _{l}$.

The difficulties of the UC model become apparent when the free energy is
minimized with respect to the free parameters of the density profile. There
are two cases depending on whether $R$ or $\lambda $ is larger. Minimizing
under the assumption $0<R<\lambda $ gives 
\begin{eqnarray}
R &=&\frac{4\gamma \left( \rho _{l}-\rho _{0}\right) ^{2}}{f\left( \rho
_{0}\right) -f\left( \rho _{l}\right) -f^{\prime }\left( \rho _{0}\right)
\left( \rho _{0}-\rho _{l}\right) } \\
R^{3}\rho _{0} &=&\frac{3}{4\pi }N-\left( \lambda ^{3}-R^{3}\right) \rho _{l}
\nonumber
\end{eqnarray}%
Now, consider what happens when $\lambda $ becomes large and $N$ is held
fixed. Since $\rho _{0}>0$, it follows that $R^{3}>\lambda ^{3}-\frac{3}{%
4\pi \rho _{l}}N$ which means that $R\sim \lambda $. This in turn implies
that $\rho _{0}<\frac{3}{4\pi R^{3}}N\sim \frac{3}{4\pi \lambda ^{3}}N$ so
that the central density tends to zero. However, at small $\rho _{0}$, one
has that $f\left( \rho _{0}\right) =\rho _{0}\ln \rho _{0}-\rho _{0}+O\left(
\rho _{0}^{2}\right) $ giving%
\begin{equation}
R=\frac{4\gamma \rho _{l}^{2}+O\left( \rho _{0}\right) }{\rho _{l}\ln \rho
_{0}-f\left( \rho _{l}\right) +O\left( \rho _{0}\right) }
\end{equation}%
and this is clearly negative for sufficiently small $\rho _{0}$, which is to
say for sufficiently large $\lambda $. Since the radius must be positive,
this is unphysical. The alternative, $R>\lambda $, leads to%
\begin{eqnarray}
R &=&-\frac{2\gamma \left( \rho _{l}-\rho _{0}\right) ^{2}}{f\left( \rho
_{0}\right) -f\left( \rho _{l}\right) -\mu \left( \rho _{0}-\rho _{l}\right) 
} \\
\rho _{0} &=&\frac{3}{4\pi \lambda ^{3}}N \nonumber
\end{eqnarray}%
Again, the central density tends to zero as $\lambda $ becomes large. In
this case, one has that 
\begin{equation}
\lim_{\rho_{0} \rightarrow 0}R=-\frac{2\gamma \rho _{l}{}^{2}}
{ -f\left( \rho _{l}\right)+\mu \rho _{l}}=-\frac{2\gamma \rho _{l}{}^{2}}
{ \beta P\left( \rho_{l} \right)}
\end{equation}%
where $P(\rho_l)$ is the pressure of the bulk liquid. It is evident that the radius again becomes negative
thus violating the assumption that $R>\lambda $. There is therefore no nontrivial
solution for large $\lambda $, which is exactly the conclusion of UC.
However, in this case the underlying free energy is clearly well defined for
all bubbles (i.e. for all values of $\rho _{0}$ and $R$) so that the lack of
solution is a statement about the structural constraint and tells us nothing
about the free energy of bubbles.

It might seem that the pathology of this model is due to the extremely
simplified density profile. However, a look at the proposed structural
constraint shows that this is not necessarily the case. In fact,
minimization of the free energy with respect to the Lagrange multiplier, $%
\alpha$, and the central density, $\rho_0$, gives $\Gamma =\frac{4\pi }{3}%
R^{3}$ and 
\begin{equation}
f^{\prime }\left( \rho_0 \right) = \mu - \frac{3 \gamma \left( \rho_l -
\rho_0 \right)^2}{R}.
\end{equation}%
The first relation shows a simple one-to-one relation between the parameter $%
\Gamma$ and the bubble radius while the second relation serves to fix the
central density. For large radii, that density tends to the density of the
bulk gas. For small radii, taking into account that the chemical potential
for a gas is generally negative and that $f^{\prime }(\rho_0) \sim
\log(\rho_0)$, it is clear that the central density tends to zero. There is
a third relation that results from minimizing with respect to $R$ but this
serves to fix the Lagrange multiplier and is of no particular interest.
Finally, note that these conclusions are independent of the value of $\rho_*$
provided it lies in the range $\left[ \rho _{v},\rho _{l}\right] $.

This analysis is consistent with the more detailed calculations of the previous section and shows that the lack of existence of a density profile that
minimizes the constrained free energy does not necessarily mean that the
unconstrained free energy is pathological in any way. In the present case,
the CNT free energy is a well-behaved function of the local density
(characterized by the parameters $\rho _{0}$, $\rho _{\infty }$ and $R$).

\section{Conclusions}

The MFEP for bubble formation in a superheated liquid was calculated using
three different methods. The most robust method appears to be the NEB. In
this approach, no a priori assumptions are made about the structure of
bubbles and the results were found to be robust with respect to the metric
used and the choice of parameters. The method yields a well-defined picture
of bubble formation and growth that is broadly consistent with that of
classical nucleation theory.

It is perhaps surprising that the use of parametrized profiles appears to
be almost as good as the NEB method in terms of the description of the
critical cluster as well as the calculation of the free energy as a function
of number deficit. For very small bubbles, the assumption of a sigmoidal
shape is not correct, but the free energy is relatively insensitive to this
error since, in a spherical geometry, the characteristic $r^2$ weighting in
the integrals minimizes the contribution of the structure near the origin while larger bubbles actually are sigmoidal.

The constraint method gave the worst results. In particular, it is not
robust: the use of the UC constraint only gives nontrivial results for small
bubbles whereas the alternative constraint discussed her gives stable
profiles of all sizes. In both cases, the profiles are discontinuous except
for the critical cluster thus giving a higher free energy than the other
methods. The failure of the UC constraint to stabilize large bubbles was
analyzed using a simple model and it was shown that the apparent instability 
was an artifact of the constraint method and tells nothing about the underlying
free energy surface. 

The conclusion, based on the combination of numerical calculations and analysis of a simple model presented here, 
is that bubble nucleation and growth does not involve an
``activated instability'' as previously claimed\cite{Corti_PRL} but, rather,
appears to follow the picture, if not the details, of classical nucleation theory rather well.

\acknowledgments
I am grateful to Gr{\'e}goire Nicolis for encouragement and for several
useful comments on this work. This work was supported in part by the
European Space Agency under contract number ESA AO-2004-070.

\bigskip 
\bibliographystyle{eplbib}
\bibliography{paper}

\end{document}